\documentclass[epj]{svjour}

\usepackage{graphicx}
\usepackage{epsfig}
\usepackage{subfigure}
\usepackage{fancyhdr}
\usepackage{xspace}
\def\makeheadbox{{
 }}
\def\mytitle{My title} 
\def\myauthors{My name}  
\def\mytype{My type of session}
\def\mysession{My session}

\def\mytitle{CDF Searches for New Physics with Photons} 
\def\myauthors{Andrey Loginov}    
\def\mytype{Contributed Talk}    
\def\mysession{Colliders - SUSY Phenomenology}

\def \Et {{E}_{\rm T}}

\newcommand{\pt}{{p}_{\rm T}}

\newcommand{\met}{\mbox{${\rm \not\! E}_{\rm T}$}}
\newcommand{\Met}{\mbox{${\rm \not\! E}_{\rm T}$}}

\def \Ht {{H}_{\rm T}}

\def\stop{\tilde{t}}

\newcommand{\Zgstar}{Z^0\kern -0.25em/\kern -0.15em\gamma^*}
\def\Zgamma{\Z\kern -0.1em/\kern -0.1em\gamma}

\def\eeggmet{ee\gamma\gamma\Met}
\def\llgg{\ell\ell\gamma\gamma}

\def\llggmet{\ell\ell\gamma\gamma\Met}
\newcommand{\lgX}{\ell\gamma+X}

\def\lgmet{\ell\gamma\Met}
\def\lgmetb{\ell\gamma\Met b}

\def\llg{\ell\ell\gamma}

\def\lgg{\ell\gamma\gamma}
\def\ggX{\gamma\gamma+X}

\def\ggg{\gamma\gamma\gamma}

\def\ll{\ell\ell}
\def\egmet{e\gamma\Met}

\def\eeggmet{ee\gamma\gamma\met}

\def\eeg{ee\gamma}

\def\mug{\mu\gamma}

\def\mumug{\mu\mu\gamma}

\def\gg{\gamma\gamma}
\def\Wg{W\gamma}

\def\lt{<}

\def\GeVc2{GeV/{c^2}}

\newcommand{\invpb}{\rm pb^{-1}}

\newcommand{\invfb}{\rm fb^{-1}}

\newcommand{\ppbar}{p{\bar p}} 

\def\ttbarg{t\bar{t}\gamma}
\def\ttg{t\bar{t}\gamma}

\newcommand{\goes}{\kern -0.18em\rightarrow\kern -0.18em}
\newcommand{\plus}{\kern -0.18em +\kern -0.18em}
\def\Z{Z^0}

\def\95cl{95 \%~C.L.}
\def\95CL{95 \%~C.L.}
\def\r#1 {$^{#1}$}

\newcommand{\Zee}{\mbox{$Z^0 \rightarrow e^+e^-$}}

\newcounter{myquestion}

\newcommand{\NONE}{\mbox{$\widetilde{\chi}_1^0$}}
\newcommand{\NTWO}{\mbox{$\widetilde{\chi}_2^0$}}

\newcommand{\none}{\NONE}
\newcommand{\ntwo}{\NTWO}

\newcommand{\GeV}{\ensuremath{\mathrm{Ge\kern -0.1em V}}\xspace}
\newcommand{\TeV}{\ensuremath{\mathrm{Te\kern -0.1em V}}\xspace}
\newcommand{\bfTeV}{\ensuremath{\bf{Te\kern -0.1em V}}\xspace}
\newcommand{\rrr}{\rightarrow}

\newcommand{\Etgamma}{\ensuremath{{E_T^{\gamma}}}}

\begin{document}
\title{CDF Searches for New Physics with Photons}
\subtitle{}
\author{A.~Loginov
}                     
%
%
\institute{Yale University, New Haven, Connecticut 06520}
%
\date{}
\abstract{
We present results of searches for new physics in final states with
photons at CDF in approximately 1$\invfb$ of $\ppbar$ collisions at
1.96 $\TeV$. We give an overview of the data-driven photon background
estimation techniques used for the analyses. We report on a search for
diphoton peaks and signature-based searches for $\ggX$, $\lgX$ ($X$ =
$e$, $\mu$, $\met$, $\gamma$) and $\ttg$.
} 
\maketitle
\section{Introduction}
\label{intro}

A crucial test of the standard model (SM) of particle
physics~\cite{SM} is to measure and understand the properties of the
highest momentum-transfer particle collisions, and therefore to study
interactions at the shortest distances. The major predictions of the
SM for these collisions are the rates for the events of a given type,
and their associated kinematic distributions.

Since the predicted high energy behavior of the SM becomes unphysical
at an interaction energy on the order of several TeV, new physical
phenomena are required. These unknown phenomena may involve new
elementary particles, new fundamental forces, and/or a modification of
space-time geometry. An anomalous production rate of a
combination of the known fundamental particles is likely to be a
manifestation of the new phenomena.

The unknown nature of possible new phenomena
(Table~\ref{models.table}) in the energy range accessible at the
Tevatron is the motivation for a ``signature-based'' search strategy
that does not focus on specific models of new
physics, but casts a wide net for new phenomena.


In the Tevatron Run I, the rare $\eeggmet$ candidate event and the
measured event rate for the signature $\ell+\gamma+\met$, which was
2.7 sigma above the SM predictions, sparked signature-based searches
in the $\gamma\gamma+X$~\cite{Toback_PRL} and
$\ell\gamma+X$~\cite{Jeff_PRL} channels. Now in Run II we have
performed these searches with an order-of-magnitude larger data set, a
higher $\ppbar$ collision energy, and the CDF II
detector~\cite{CDFII}. In these proceedings we report on a search for
diphoton peaks and signature-based searches for $\ggX$, $\lgX$ ($X$ =
$e$, $\mu$, $\met$, $\gamma$) and $\ttg$.

\begin{table}[!t]
\begin{center}
\begin{tabular}{p{3cm}p{4cm}}
\hline
\multicolumn{2}{l}{SUSY}\\

${\tilde{\ell}} \rrr \ntwo + \ell$, &
$\gg\met$
{$\lgmet$}, $\ell$ = $e$ or $\mu$ \\

$\ntwo\rrr\gamma \none$, $\none\rrr\gamma \tilde{G}$ &

{ $\gamma \ll\met$,$\gg \ll\met$}, $\ell$ = $e$ or $\mu$ \\

\hline
\multicolumn{2}{l}{Technicolor}\\

$\omega_T, \rho_T \rrr \gamma\pi_T$ &
{$\ttbarg$, $\ggg$} \\

$\pi_T \rrr \gg$ &
$\ll\gg$, $\ll\gg\met$, $\ell$ = $e$ or $\mu$ \\

\hline
\multicolumn{2}{l}{Compositeness}\\

$X^*\rrr\gamma X$ &
$\llg$, $\llgg$, $\ell$ = $e$ or $\mu$ \\

\hline
\multicolumn{2}{l}{Extra Dimensions (RS)}\\

$G\rrr\gg$ &
$\gg$ \\

\hline
\end{tabular}

\caption
{Some of the models beyond the SM which predict signatures
presented in the talk.}
\label{models.table}
\end{center}
\end{table}


\section{Techniques}
\label{techniques} 



High $\pt$ photons are copiously produced by decays of hadrons in jets
resulting from a scattered quark or gluon. In particular, mesons such
as the $\pi^0$ or $\eta$ decay to photons, which may satisfy the
photon selection criteria (see Sec.~\ref{isolation}). Another
important source of non-prompt (``fake'') photons is an electron that
undergoes hard bremsstrahlung or whose track is lost due to tracking
inefficiency and is misidentified as a photon (see Sec.~\ref{egamma}).

\subsection{Isolation}
\label{isolation} 

\begin{figure}[!t]
\vspace*{-0.2in}
  \begin{center}
    \mbox{
\subfigure[Tracking isolation]
{\epsfig{file=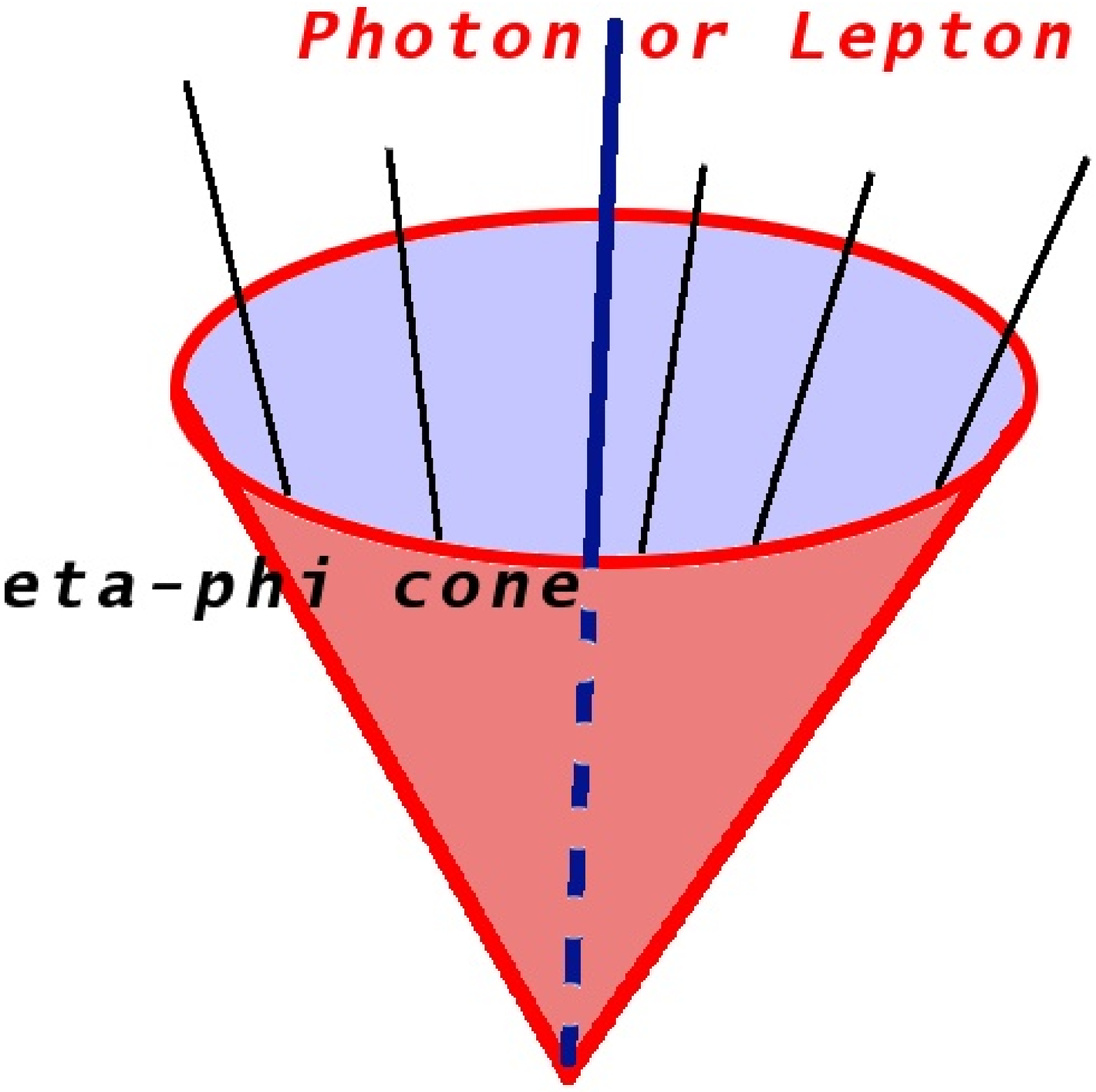,width=3.75cm, angle=0,clip=}} \quad
\hskip-0.1in
\subfigure[Calorimeter isolation]
{\epsfig{file=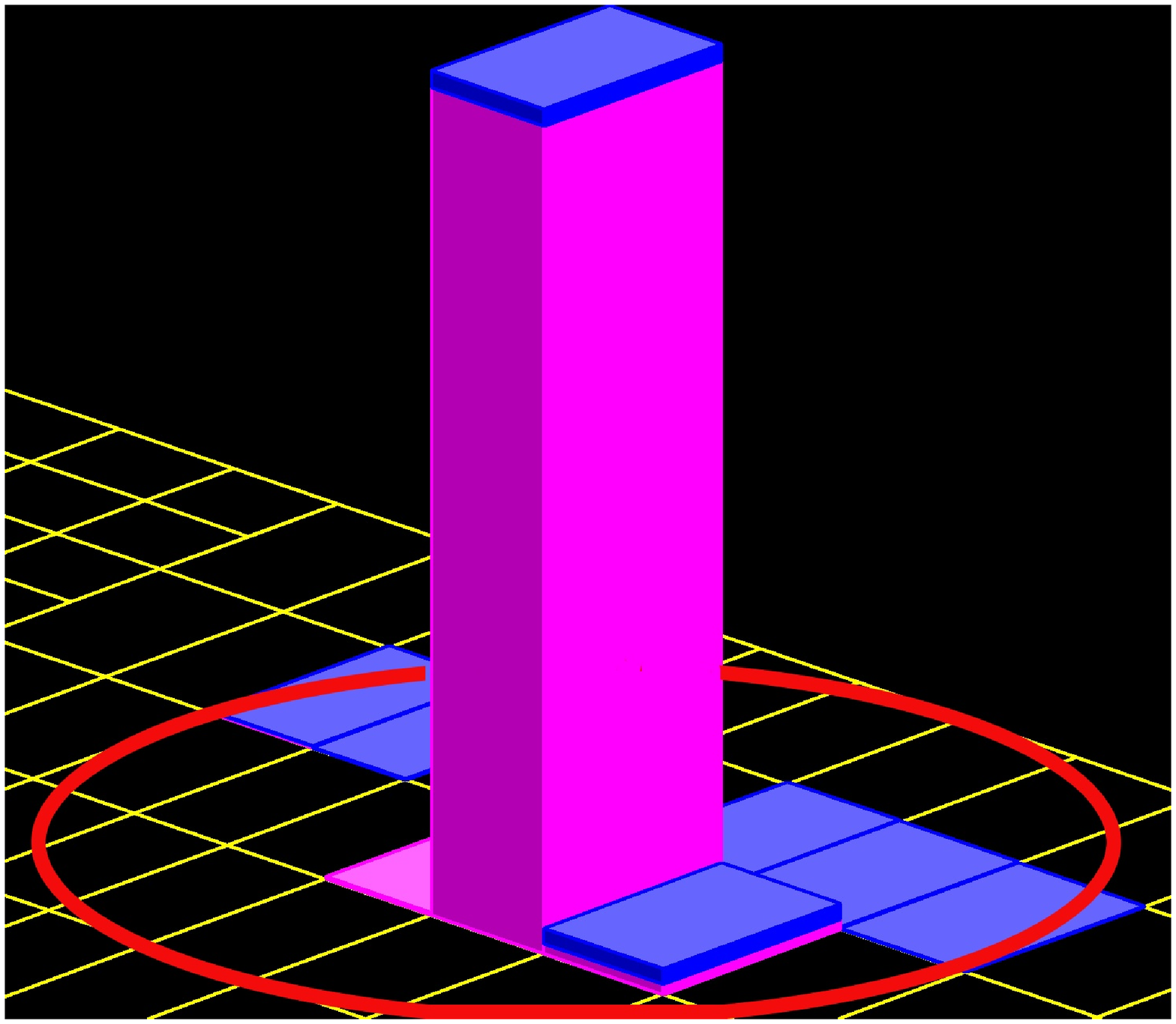,width=4.2cm,angle=0, clip=}} \quad
}
\vskip-0.2in
\caption
{Tracking isolation is the $\sum\pt$ of extra tracks around seed track
in an $\eta-\varphi$ cone. Calorimeter isolation is the $\sum\Et$ of extra
towers around the seed towers in an $\eta-\varphi$ cone. Isolation differs for
prompt photons and for 'fake' ones.}
\label{isolation_scatch.figure}
\end{center}
\end{figure}

The backgrounds due to events with jets misidentified as photons or
leptons can be estimated by studying the total $\Et$ or $\pt$ of
towers or tracks in a cone in $\eta-\varphi$ space of radius R=0.4
around the photon cluster or lepton track (see
Fig.~\ref{isolation_scatch.figure}). For instance, we estimate the
jet-faking-photon background (``$j\rrr\gamma$'') for $\llg$ and
$\lgmet$ searches~\cite{Loginov_PRD} by measuring energy in the
calorimeter nearby the photon candidate.


\begin{figure}[!b]
\vspace*{-0.2in}
  \begin{center}
    \mbox{
\subfigure[Background estimate for $\egmet$]
{\epsfig{file=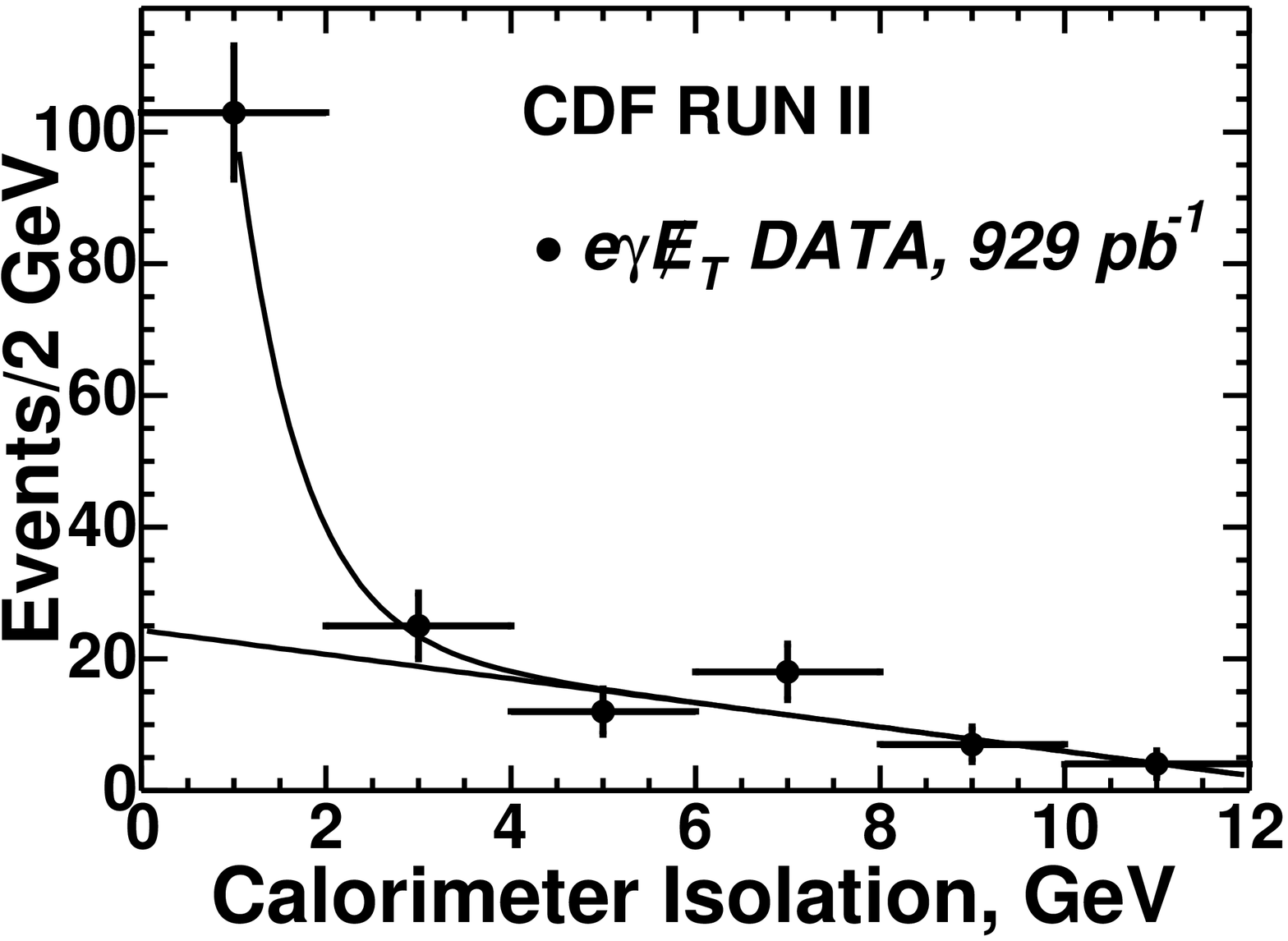,width=3.7cm, angle=0,clip=}} \quad
}
\hskip-0.1in
\subfigure[Calorimeter isolation for ``$\pi^0$'']
{\epsfig{file=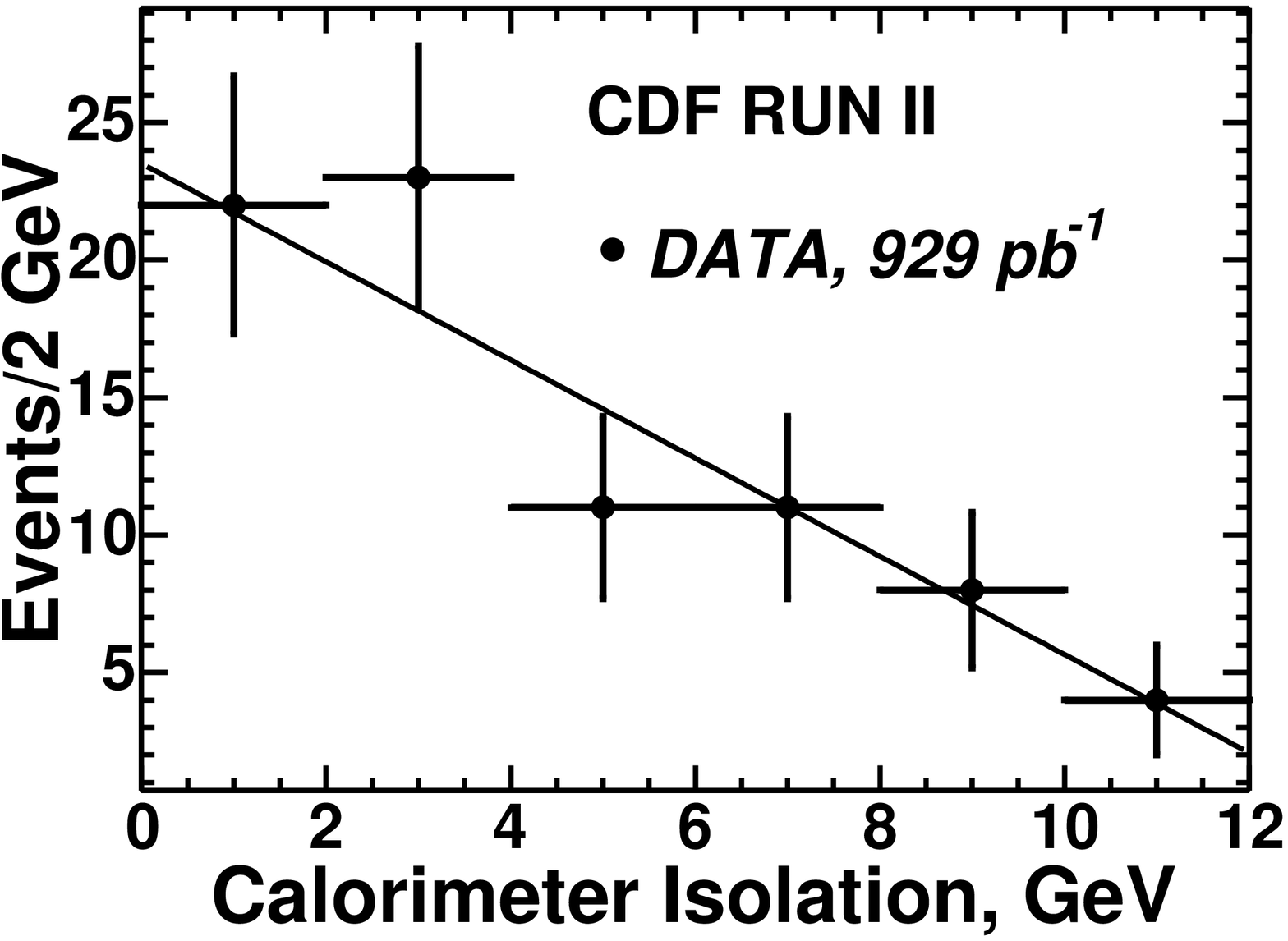,width=3.7cm,angle=0, clip=}} \quad
\vskip-0.2in
\caption{
The method and data used to estimate the number of background events
from jets misidentified as photons for the $\egmet$ sample (a). The number
of events is plotted versus the total (electromagnetic plus hadronic)
calorimeter energy, $\Et^{Iso}$, in a cone in $\eta$-$\varphi$ space
around the photon. This distribution is fit to the shape
measured for electrons from $\Zee$ decays plus a linear background,
obtained from a fake photon (``$\pi^0$'' sample) (b).}
\label{egmet_isolation.figure}
\end{center}
\end{figure}

Figure~\ref{egmet_isolation.figure} (a) shows the distribution in the
total (electromagnetic plus hadronic) calorimeter energy, $\Et^{Iso}$,
in a cone of radius $R=0.4$ in $\eta$-$\varphi$ space around the
photon candidate for the $\egmet$ sample. This distribution is fit to
the shape measured for electrons from $\Zee$ decays plus a linear
background.

To verify the linear behavior of the background we select a sample of
``fake photons'' by requiring the photon candidate to fail the cluster
profile criteria. In addition we do not apply the calorimeter and
track isolation requirements. The distribution of the total
calorimeter energy, $\Et^{Iso}$, in a cone of radius $R=0.4$ in
$\eta$-$\varphi$ space around the fake photon candidate is shown in
Fig.~\ref{egmet_isolation.figure} (b).

\subsection{Phoenix Tracking}
\label{egamma} 

If an electron undergoes hard photon bremsstrahlung or no electron
track is reconstructed due to the tracking inefficiency, we still can
look for track segments in the silicon tracker. To find a segment we
seed a ``Phoenix'' track from an electromagnetic calorimeter cluster
and an event vertex and search for hits along the expected arc.


\begin{figure}[!h]
\vspace*{-0.2in}
  \begin{center}
    \mbox{
\subfigure[Method]
{\epsfig{file=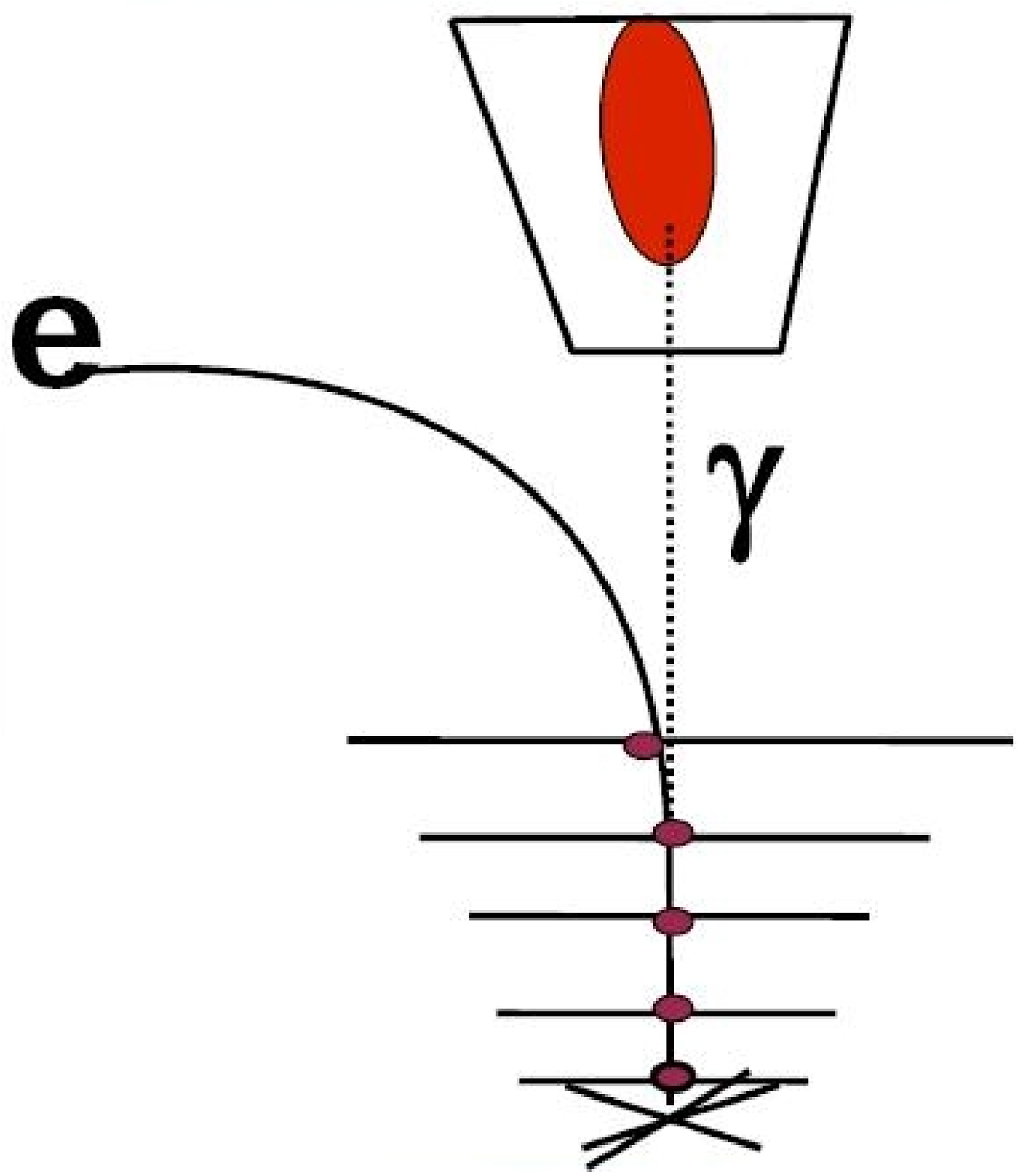,width=2.7cm, angle=0,clip=}} \quad
}
\hskip-0.1in
\subfigure[$P^{e}_{\gamma}$]
{\epsfig{file=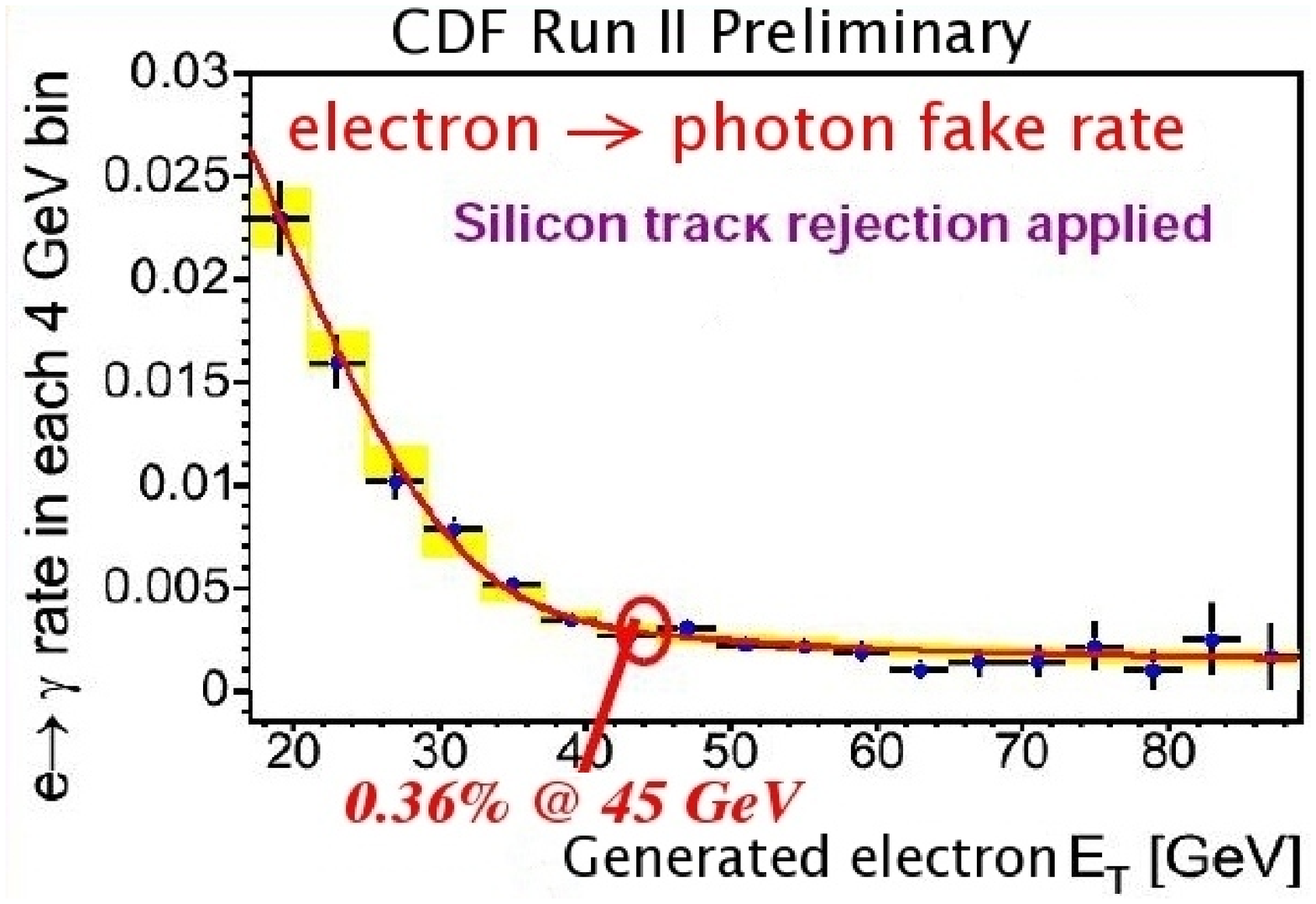,width=5.1cm,angle=0, clip=}} \quad
\vskip-0.2in
\caption{
The method and data used to estimate the probability that an electron
undergoes hard photon bremsstrahlung and is misidentified as a photon,
$P^{e}_{\gamma}$.}
\label{egamma.figure}
\end{center}
\end{figure}

The probability that an electron undergoes hard photon bremsstrahlung
and is misidentified as a photon (``$e\rrr\gamma$''),
$P^{e}_{\gamma}$, is measured from the control subsample of
back-to-back $e\gamma$ events consistent with originating from
$Z^{0}\rightarrow e^+e^-$ production (Fig.~\ref{egamma.figure}). The $\Et$
dependence of $P_{e\rrr\gamma}$ is obtained from $\Zee$ simulated
sample, and then normalized to data.

\section{Searches}
\label{searches}

\subsection{Search for Diphoton Peaks}
\label{diphoton_peaks}

One of the signatures for new, heavy particles is a narrow mass
resonance decaying to $\gg$. This signature could arise from extra
spatial dimensions, as in the Randall-Sundrum (RS)
model~\cite{rs}. The spin-2 nature of the graviton, decaying by either
$s$- or $p$-wave states, favors searching in the $\gg$ channel, where
the branching ratio is twice that of any single dilepton channel.

In the RS model the widths and masses of the resonances are dependent
on the parameter $k/\overline{M}_{Pl}$, where $\overline{M}_{Pl}$ is
the effective four-dimensional (reduced) Planck scale and $k$ is a
curvature parameter.



\begin{figure}[!t]
\vspace*{-0.2in}
\begin{center}
\includegraphics[width=1.0\linewidth,clip=]{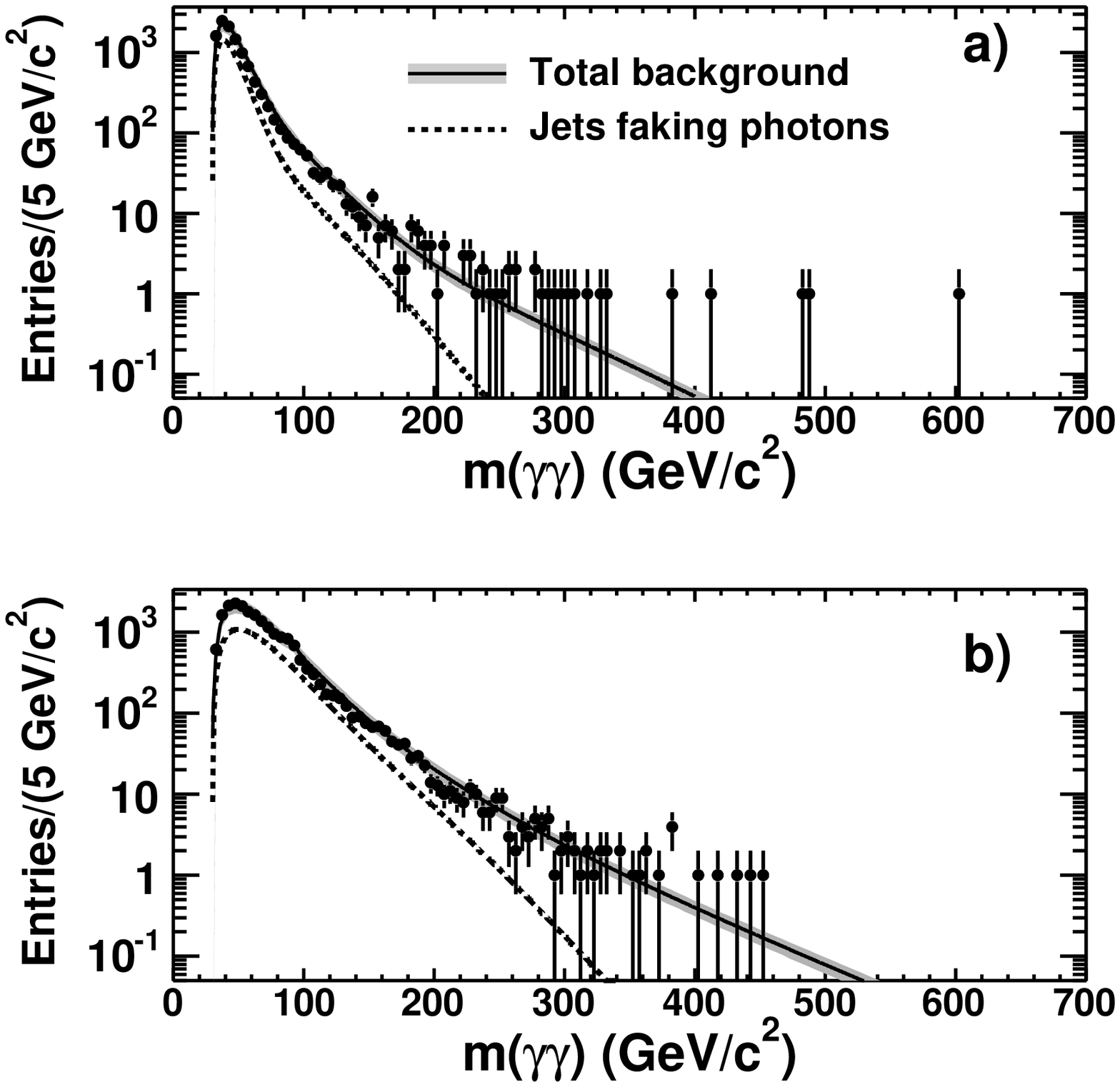}
\caption{
The mass distribution in the CC (a) and CP (b) signal regions
with the background overlaid. The points are the data. 
The dotted line shows the jets which fake photons 
as predicted from the photon-like jet sample, and the solid line shows
this background plus the SM $\gg$.}
\label{Fig:display_signal}
\end{center}
\end{figure}

There are two significant components to the diphoton data sample: SM
$\gg$ production and hadronic jets faking photons. The diphoton mass
distribution of the jet background is derived from a sample of
photon-like jets obtained by loosening the photon selection criteria,
removing events which pass all the signal selection
requirements. Figure~\ref{Fig:display_signal} shows the observed mass
spectra compared to the predictions.

\begin{figure}[!b]
\vspace*{-0.2in}
\begin{center}
\includegraphics[width=1.0\linewidth,clip=]{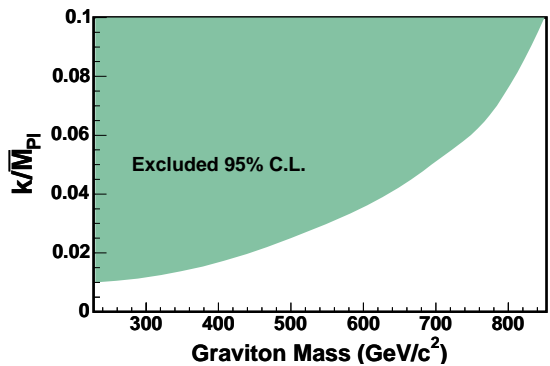}
\caption{The 95\% C.L. excluded region 
in the plane of $k/\overline{M}_{Pl}$ versus graviton mass. 
}
\label{Fig:exclude_region}
\end{center}
\end{figure}

The values of $k$ must be large enough to be consistent with the
apparent weakness of gravity, but small enough to prevent the theory
from becoming non-per\-turbative. Therefore, we examine $0.01 <
k/\overline{M}_{Pl} < 0.1$.


To parametrize the background for setting limits, we fit the mass
spectra to the combination of a polynomial multiplied by the sum of
two exponentials and the expected SM diphoton shape. The 95\%
confidence-level (C.L.) excluded region in the $k/\overline{M}_{Pl}$
versus graviton-mass plane is shown in Fig.~\ref{Fig:exclude_region}.


To conclude, we find no evidence of new physics in the diphoton
distribution. We evaluate 95\% C.L. limits in one model of
hypothetical new diphoton production and exclude RS gravitons below
masses ranging from 230~GeV/$c^{2}$ to 850~GeV$/c^{2}$, for a coupling
parameter $k/\overline{M}_{Pl}$ of 0.01 to 0.1.  This results in a
significant improvement, at high mass, over the previous best
available limit. Details on this analysis are available in
Ref.~\cite{gg_rs_PRL}.

\subsection{Search for $\ggX$}
\label{ggx}

%

For the $\ggX$ search we require two photons in the central region
with $\Etgamma>13~\GeV$ in an event. We then perform inclusive
searches for additional objects, such as $\met$, another $\gamma$, $e$
or $\mu$.

\begin{figure}[!t]
\vspace*{-0.2in}
\begin{center}
\hspace*{-0.1in}
\includegraphics*[angle=0,width=0.4\textwidth]{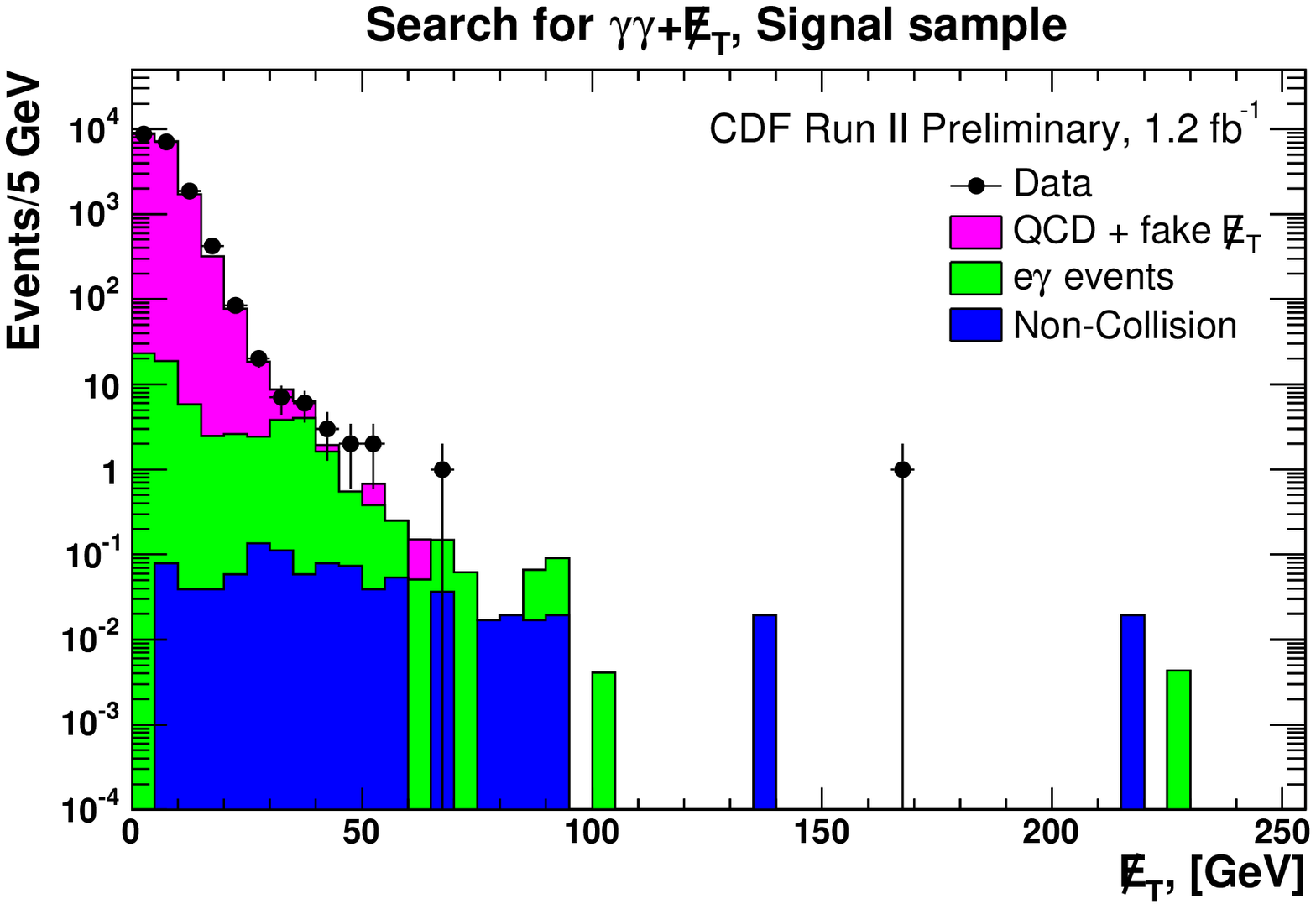}
\caption
{The distribution for the $\met$ in $\gg$ events.}
\label{gg_met.figure}
\end{center}
\end{figure}

The distribution of the $\met$ in $\gg$ events is shown in
Fig.~\ref{gg_met.figure}. At low $\met$ the background is dominated by
$\gg$, $j\gamma$ and $jj$ ($j\rrr\gamma$) events with mismeasured
$\met$. Intermediate and large $\met$ regions are populated by
$\Wg\rrr e\nu\gamma$ ($e\rrr\gamma$) and electromagnetic showers due to
cosmic-ray and beam halo muons. The overall agreement between data and
the prediction is good. For $\met>50~\GeV$ we find 4 $\gg\met$ events
compared to an expectation of $1.6 \pm 0.3$ events.

\begin{figure}[!b]
\vspace*{-0.1in}
\begin{center}
\hspace*{-0.1in}
\includegraphics*[angle=0,width=0.3\textwidth]{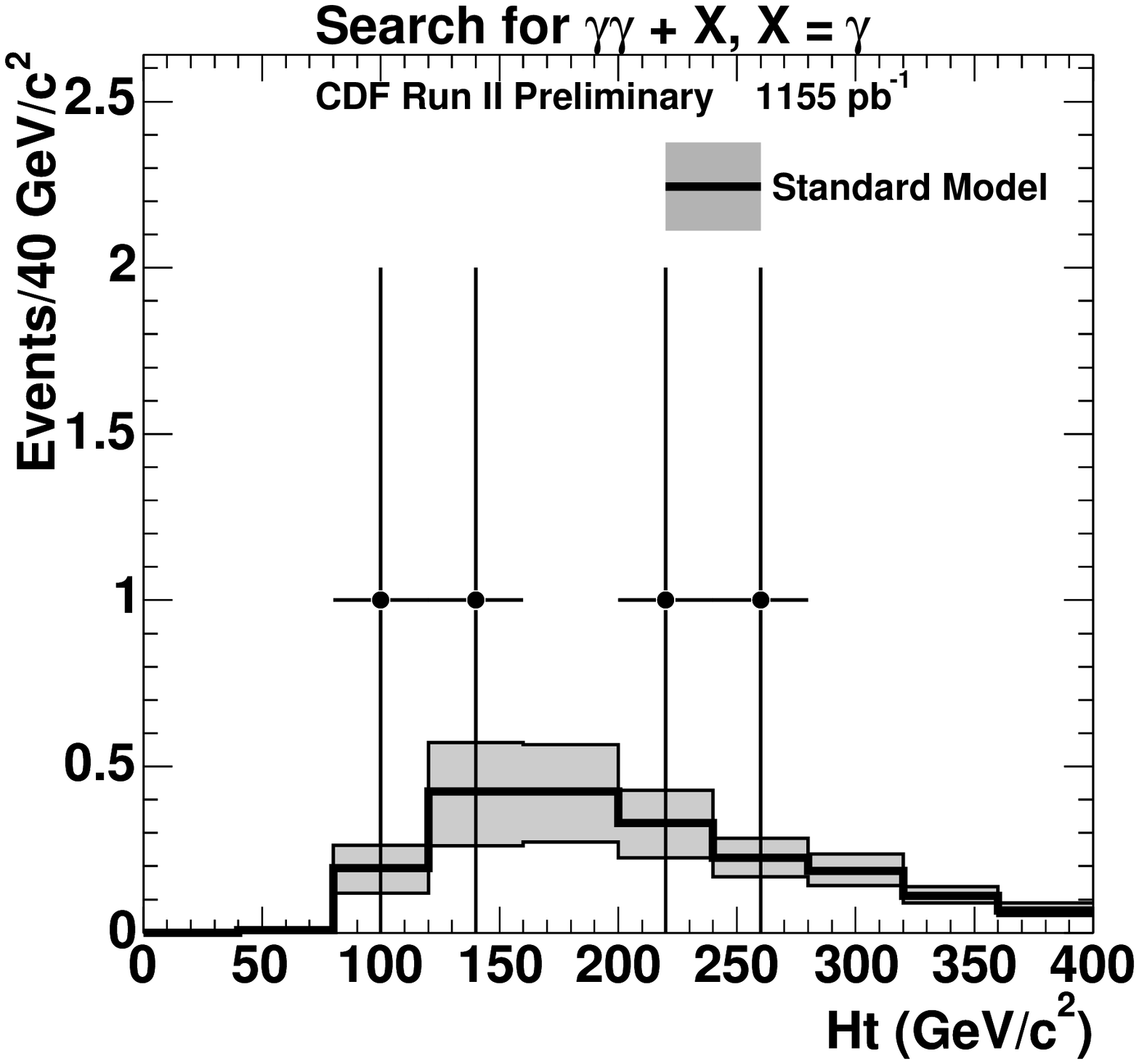}
\caption
{$\Ht$ distribution in $\ggg$ candidate events, where $\Ht$ is the sum
of the transverse energies of the photons, jets and $\met$.}
\label{ggg_ht.figure}
\end{center}
\end{figure}

For the $\ggg$ search we require an additional central photon with
$\Etgamma>13~\GeV$. SM triphoton production is estimated to be $0.80
\pm 0.15$ events. Fake-photon background is estimated to be $1.4 \pm
0.6$ events. We find 4 $\ggg$ events compared to an expectation of
$2.2 \pm 0.6$. The distribution for $\Ht$, the sum of the transverse
energies of the photons, jets and $\met$, is shown in
Fig.~\ref{ggg_ht.figure}. There is good agreement with the expected SM
prediction.

\subsection{Search for $\lgX$}
\label{lgx}

\begin{figure}[!t]
\vspace*{-0.2in}
\begin{center}
\hspace*{-0.1in}
\includegraphics*[angle =90,width=0.50\textwidth]{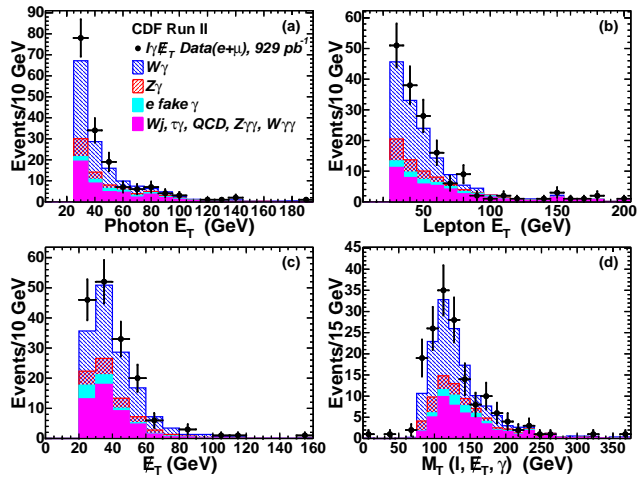}
\caption{ The distributions for events in the $\lgmet$ sample
  (points) in a) the $\Et$ of the photon; b) the $\Et$ of the lepton
  ($e$ or $\mu$); c) the missing transverse energy, $\met$; and d) the
  transverse mass of the $\lgmet$ system.  The histograms show the
  expected SM contributions, including estimated backgrounds from
  misidentified photons and leptons.}
\label{wg_fig1_leptons}
\end{center}
\end{figure}


To test whether either the $\llggmet$ or $\lgmet$ Run~I results
included new physics effects, we have repeated the $\lgX$ search in
929 $\invpb$ of Run~II data. Using the same selection criteria makes
this search an {\it a priori} test, as opposed to the Run I
measurement.

\begin{figure}[!b]
\vspace*{-0.2in}
\begin{center}
\includegraphics*[width=0.18\textwidth, angle=90,clip=]{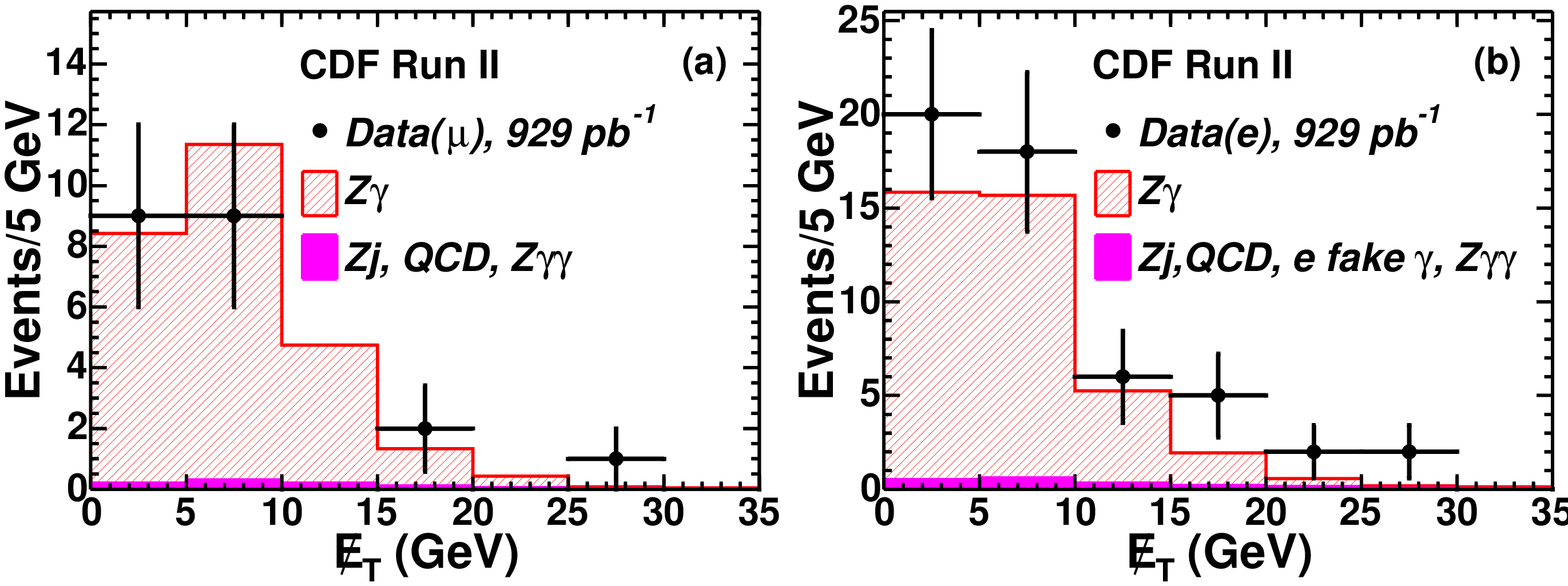}
\caption {The $\met$ distributions observed in the inclusive search for a)
$\mumug$ events and b) $\eeg$ events. The histograms show the expected
SM contributions.}
\label{zg_fig3_leptons}
\end{center}
\end{figure}



We find 163 $\lgmet+X$ events, compared to an expectation of $150.6
\pm 13.0$ events. The predicted and observed kinematic distributions
for the $\lgmet+X$ signature are compared in
Fig.~\ref{wg_fig1_leptons}. 
We observe no $\lgg$ or $e\mug$ events, compared to the expectation of
$0.62 \pm 0.15$ and $1.0 \pm 0.3$ events, respectively.

We observe 74 $\llg + X$ events, compared to an expectation of $65.1
\pm 7.7$ events. The predicted and observed kinematic distributions for the
$\met$ in $\llg+X$ events are shown in Fig.~\ref{zg_fig3_leptons}. We
find 3 $\llg$ events with $\met> 25~\GeV$, compared to an expectation
of $0.6 \pm 0.1$ events (mostly SM $\llg$), corresponding to a
likelihood of 2.4\%.

We observe no $\llg$ events with multiple photons and so
find no events like the $\eeggmet$ event of Run I.

The $\eeggmet$ event thus remains a single event selected {\it a
posteriori} as interesting, but whether it was from SM
$WW\gamma\gamma$ production, a rare background, or a new physics
process, we cannot determine. Details on this analysis are
available in Ref.~\cite{Loginov_PRD}.

\subsection{Search for $\ttg$}
\label{ttg}

\begin{figure}[!t]
\vspace*{-0.2in}
\begin{center}
\hspace*{-0.1in}
\includegraphics*[angle=0,width=0.35\textwidth]{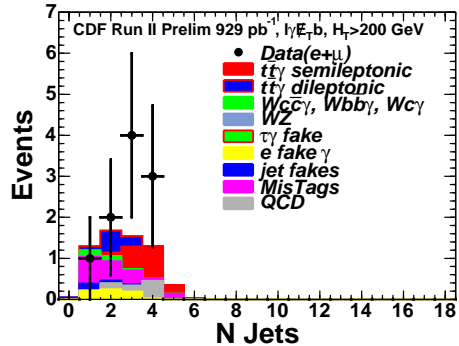}
\caption
{The distribution for the total number of jets in $\lgmetb$ events
with $\Ht>200~\GeV$, where $\Ht$ is the total transverse energy, the
sum of the transverse energies of the lepton, photon, jets and
$\met$. For the search for $\ttg$ events in addition to $\Ht>200~\GeV$
we also require $N_{jets}\geq 3$. The histogram shows the expected SM
contributions.}
\label{lgmetb_ht_njets}
\end{center}
\end{figure}

We have searched for the anomalous production of $\lgmetb$ events,
such as might be produced in the anomalous production of top quarks,
using 929 $\invpb$ of data. SM sources of such events include top
quark pairs with an additional photon, $\ttg$, as well as $\Wg$
production with $b$ or $c$ quark jets, both of which can result in the
signature of a high-$\pt$ lepton, photon, $b$-tagged jet, and
$\met$. When one in addition requires large $\Ht$, and~$\geq 3$ jets,
radiative top events dominate the SM prediction. 
The search is an
extension of a investigation of the $\lgX$ signature (Sec.~\ref{lgx}).

We find 10 $\lgmetb+\Ht$ ($\Ht>200~\GeV$) events (6 for the $e$
channel and 4 for the $\mu$ channel) versus an expectation of $7.2
\pm 1.0$ events ($4.9 \pm 0.8$ and $2.3 \pm 0.6$ for the $e$ and $\mu$ 
channels, respectively). Figure~\ref{lgmetb_ht_njets} shows the
distribution for the total number of jets in $\lgmetb$ events with
$\Ht>200~\GeV$.

For the search for $\ttg$ events we require $\Ht>200~\GeV$ along with
$N_{jets}\geq 3$. We find 7 such events (4 for the $e$ channel and 3 for
the $\mu$ channel) versus an expectation of $3.6 \pm 0.8$ events ($2.3
\pm 0.6$ and $1.3 \pm 0.5$ for the $e$ and $\mu$ channels, respectively).

We find that the numbers of events agree with SM predictions. With
enough statistics, in the context of the SM a measurement of the
radiative decay of the top quark directly measures the charge of the
quark~\cite{Baur_Orr_Rainwater}. More generally, events with a top
pair with an additional radiated boson, $\gamma$, $W$/$Z$, or Higgs,
both measure the SM couplings and provide windows with small SM
cross-sections into which to look for physics processes coupling to
the top quark outside the SM~\cite{MrennaKane}.

\section{Conclusion}

The techniques developed and experience gained will be useful for the
Tevatron and for the LHC. We find that the number of events in the
photon searches agree with the SM predictions. We find no events like
the $\eeggmet$ event of Run I. There are signatures with small number
of events expected in datasets analyzed, such as $\llg+\met$, $\ggg$
and $\ttg$. We look forward to more data in Run~II to update and
improve the sensitivity of searches for new physics in photon final
states.

\end{document}